# Controls over Spreadsheets for Financial Reporting in Practice

Nancy Coster*, Linda Leon*, Lawrence Kalbers*, and Dolphy Abraham**
*Loyola Marymount University, Los Angeles, California (USA)
**Alliance University, Bangalore, India
Nancy.Coster@lmu.edu

**ABSTRACT**

*Past studies show that only a small percent of organizations implement and enforce formal rules or informal guidelines for the designing, testing, documenting, using, modifying, sharing and archiving of spreadsheet models. Due to lack of such policies, there has been little research on how companies can effectively govern spreadsheets throughout their life cycle. This paper describes a survey involving 38 participants from the United States, representing companies that were working on compliance with the Sarbanes-Oxley Act of 2002 (SOX) as it relates to spreadsheets for financial reporting. The findings of this survey describe specific controls organizations have implemented to manage spreadsheets for financial reporting throughout the spreadsheet's lifecycle. Our findings indicate that there are problems in all stages of a spreadsheet's life cycle and suggest several important areas for future research.*

**1 INTRODUCTION**

It is broadly accepted that errors are prevalent in spreadsheets [Panko, 2006]. Spreadsheet risk can be defined as the likelihood of adverse operational or financial consequences resulting from use of a spreadsheet. To date, most spreadsheet research has focused primarily on understanding and mitigating spreadsheet risks associated with quantitative errors that occur during system development, the first part of the system's life cycle. This line of research investigates and often suggests implementation of more formal software engineering techniques during the creation of a spreadsheet [Leon, Abraham & Kalbers, 2010; Grossman & Ozluk, 2010; Panko, 2006]. While there are definite risks of developing an erroneous spreadsheet, there are additional and just as serious risks associated with the maintenance and operational use of the spreadsheet throughout the later parts of its life cycle.

We surveyed 38 companies working on compliance with the Sarbanes-Oxley Act of 2002 (SOX) as it relates to spreadsheets associated with financial reporting. We collected information about the controls and processes they have implemented in their organization as well as the difficulties/challenges they have encountered. This paper presents the findings of this survey. While these findings describe spreadsheets associated with financial reporting, the identification of effective controls and processes are applicable to other key spreadsheets in an organization and therefore should be considered in developing best practices for IT governance. Similarly we expect all organizations will encounter the same difficulties that these organizations are facing, but many non-regulated companies will avoid addressing risk until there is an easy and effective way of dealing with the problem.

This paper is outlined as follows: the next section provides an overview of spreadsheet controls and the accountability SOX introduces that motivates organizations to implement controls and processes for spreadsheet development and use. The survey methodology is then discussed and the results of the survey are described. Finally, conclusions and suggestions for practice and future research are made as we identify the areas where companies are struggling to effectively control spreadsheets.

## 2 OVERVIEW OF SPREADSHEET CONTROLS AND SOX ACCOUNTABILITY

### 2.1 Background

Panko and Halverson [1996] outlined a taxonomy of spreadsheet research issues as a three dimensional cube, where one important dimension was life cycle stage. As a first step in the outlined research, they created a separate taxonomy of development and testing error types [Panko and Halverson, 2001]. Panko and Aurigemma [2010] revised this taxonomy but noted that the lifecycle dimension was not addressed in either version: their studies did not examine a spreadsheet's ongoing use after development. In the taxonomy proposed by Rajalingham, Chadwick & Knight [2000], errors that end-users can make, such as data entry errors or interpretation errors, as well as the user's intention to create fraud, were considered. This taxonomy represented a first attempt to define and classify spreadsheet risks during operational use. Since then several governance issues have been identified which contribute to the risk of a spreadsheet after its development, such as maintenance, documentation, version control, privacy issues and separation of duties.

Basic spreadsheet programs lack the embedded logic and data controls necessary to prevent errors and misuse during operational use, so organizations need to apply manual or automated control processes to help mitigate spreadsheet risks by ensuring that appropriate tools are used to minimize, detect, and resolve errors throughout the entire life cycle. In general, end-users are resistive to attempts to control and restrict the development, sharing and use of self-generated models. The challenge is to identify effective controls that can help an organization improve the integrity of its spreadsheets without the controls being prohibitively time-consuming or expensive to implement and without interfering with the benefits of the spreadsheet medium.

Surveys show that most organizations have no formal policies to ensure the integrity of its operational spreadsheets [Panko, 1998; Caulkins, Morrison & Wiedemann, 2007; Lawson et al., 2009]. Companies reported that while informal guidelines were common, formal guidelines existed in only about half of the organizations. Neither the formal rules nor the informal guidelines were usually implemented and enforced throughout the development, testing, auditing, and modification stages of the spreadsheet life cycle, despite all of the literature on the prevalence of spreadsheet errors in organizations. One area where the corporate culture has changed is in financial reporting [Rittweger & Langan, 2010]. Sarbanes-Oxley regulations (SOX) hold publicly traded companies accountable for implementing and evaluating their spreadsheet controls for financial reporting. The Public Company Accounting Oversight Board's (PCAOB) Auditing Standard 5 identifies the need for a combination of preventive and detective controls to prevent and detect errors or fraud in financial reporting [PCAOB, 2007]. In 2004, several surveys reported that 80-95% of U.S. firms use spreadsheets for financial reporting [Panko, 2006]. Thus, SOX forces many publicly traded companies to view end-user developed spreadsheet models that impact financial reporting similar to formal information systems used for financial reporting.

## 2.2 Background of SOX and Controls

In the U.S., as a result of various financial frauds and scandals over the past two decades, the Sarbanes-Oxley Act of 2002 (SOX) [U.S. Congress, 2002] initiated new policies, procedures, and disclosures in financial reporting for publicly held companies. As a result, when external audit firms identify material weaknesses in a company's financial reporting process a description of the weakness or deficiency is documented in the company's annual 10-K report. Audit Analytics is a public company intelligence service that provides detailed research on over 20,000 public companies. Based on the companies included in their database, there were 113 10-Ks that recorded material weaknesses as the result of inadequate spreadsheet controls for 77 different companies between 2004 and the first half of 2008 [Leon, Abraham & Kalbers, 2010]. For example, in 2006, Design Within Reach Inc. was identified as having the following material weakness: "Specifically, controls were not designed and in place throughout the year to ensure that access was restricted to appropriate personnel and that unauthorized modification of the data or formulas within spreadsheets was prevented" [Design Within Reach Inc. 10-K, 2006].

The external audit firms have provided documented guidance that no one in the organization is assuming accountability for spreadsheet risk management and control deficiencies [Protiviti Inc., 2008]. Ultimately though, senior management is the party that will be held accountable for the identified deficiencies. Therefore, senior executives should communicate an end-user computing policy to define the spreadsheet risk management requirements expected from the organization [PricewaterhouseCoopers, 2004]. This policy must define effective processes and enact appropriate monitoring to ensure compliance with these processes. From this policy, an operating model defining accountability, roles and responsibilities, processes, controls, and control standards can be created [O'Beirne, 2005]. Finally, the company should document the usage of the controls and processes outlined in the operating model.

It is advisable for companies to adopt a framework as a foundation for developing policies and procedures for spreadsheet controls. Many companies and auditors have adopted Control Objectives for Information and Related Technology (CobiT) [IT Governance Institute, 2007] to address IT compliance for SOX [Blum, 2005]. Other useful guidance for the development and assessment of spreadsheets also exists. The Institute of Internal Auditors recently issued a practice guide for user-developed applications (UDAs), which includes guidelines for controlling and auditing UDAs using a risk-based assessment of financial, operational, and compliance materiality [Institute of Internal Auditors, 2010]. PricewaterhouseCoopers [2004] proposes that organizations use a high-level five step process to manage spreadsheet risk:
1. Create an inventory of spreadsheets that are in the scope of SOX regulations
2. Perform a risk assessment of financial misstatement (materiality and likelihood) by evaluating the use and complexity of the spreadsheet
3. Determine the necessary level of controls for "key" spreadsheets
4. Evaluate existing controls for each spreadsheet
5. Develop action plans for remediating control deficiencies

General types of controls that can be considered include change controls, version controls, access controls, input controls, security and integrity of data, documentation, development lifecycle, back-ups, archiving, logic inspection, segregation of duties, and overall analytics [PricewaterhouseCoopers, 2004]. The accountability that SOX imposes makes it critical for companies to consider how these different types of controls should be implemented in their operations, which includes defining who should be responsible for their implementation and for monitoring their effectiveness. It is critical that an organization clearly define the roles and responsibilities of different organizational stakeholders, which includes developers, business users, business owners (who are defined as the people responsible for having the spreadsheet developed), IT and IS security officers, independent review groups, the accounting department, and internal auditors. It is often the case however that one person performs several roles, such as the business owner, developer, user and reviewer.

## 3 SURVEY METHOD AND SAMPLE

We conducted an online survey of 38 U.S. publicly traded companies to study how organizations define the roles and responsibilities of different stakeholders for various types of controls. In particular, we investigated what these companies were doing to comply with SOX, what roles various stakeholders played, which stakeholders were responsible for ensuring processes were implemented and which processes organizations found most challenging to control. The survey included items related to the seven-stage model of spreadsheets: designing, testing, documenting, using, modifying, sharing, and archiving [Lawson et al., 2009]. The survey questions were designed to elicit responses about material and/or critical spreadsheet applications used in the financial reporting process, where "material" and/or "critical" spreadsheet applications were defined as being significant to the financial statements and/or footnotes, and probably identified as "in scope" for purposes of SOX compliance. A longer and a shorter version of the survey were developed. This was done to increase participation. The short survey provided an option to continue and complete the long survey.

We had several objectives in selecting our sample from the population of U.S. public companies. First, we wanted a sample of public companies that varied in size and industry. This was intended to provide a better opportunity to generalize to all public companies. Second, we desired a person from each participating company to complete the survey who had the necessary knowledge of SOX compliance and spreadsheet controls in their organization. This was intended to increase the reliability of the responses.

Several approaches were taken to elicit responses to meet our objectives. First, participants were sought by posting a link to the electronic survey on local LinkedIn groups associated with the Information Systems Audit and Control Association and the Institute of Internal Auditors. Second, a list of the 200 largest public companies in Southern California and a random sample from the S&P 1500 were called in an attempt to identify the contact name of a qualified individual in the organization. After obtaining a specific name and email address, an email was sent to each contact who had expressed a willingness to participate. Last, some personal contacts of the authors were emailed. The emails opened with the sentence, "You have been identified as the person who is most knowledgeable about how spreadsheets associated with financial reporting for SOX compliance are being managed in your company" to emphasize the need for an appropriate person to complete the survey. The email also briefly described the study and provided a link to the online surveys. All participants were given a choice between the shorter and longer versions of the anonymous survey.

There were a total of 38 respondents to our online surveys—26 responded to the longer survey and 12 to the shorter version. The total sample population cannot be estimated due to the various approaches taken, therefore no response rate can be calculated. Table 1 presents the 15 different industries represented in the sample. Manufacturing and entertainment had the largest number of respondents, representing 23.7% and 15.8% of the sample, respectively. The size of the companies responding, measured in assets, ranged from less than $24 million to over $100 billion, with about 74% of companies with assets greater than $1 billion (see Table 2). Of the 25 respondents that provided their job title, the majority (16) were associated with internal audit, followed by SOX or General Compliance positions (5). In addition, 15 respondents were also the person responsible for SOX Compliance and/or spreadsheet controls in the organization. These results indicate that respondents were at an appropriate level of the organization and knowledgeable about controls over spreadsheets for purposes of this study.

## 4 FINDINGS

In this section we report the results of the survey. There were 26 respondents that answered the longer survey version, and 12 that answered the shorter version. Therefore, there was a maximum of 26 or 38 responses for items from the longer survey and the total responses, respectively. In order to increase participation, participants were allowed to skip questions that they did not feel comfortable answering. Thus the responses for some items are lower than the overall number of participants for the survey. Results are presented in percentages and/or numbers. Care is taken to show the total responses in each case.

Table 1 Industries Represented in the Sample

|  | Response Percent | Response Count |
|---|---|---|
| Aerospace & Defense | 5.3% | 2 |
| Construction | 2.6% | 1 |
| Engineering & Related Services | 2.6% | 1 |
| Entertainment | 15.8% | 6 |
| Financial Services | 7.9% | 3 |
| Health Care | 5.3% | 2 |
| Insurance | 2.6% | 1 |
| Manufacturing | 23.7% | 9 |
| Real Estate | 2.6% | 1 |
| Restaurants | 5.3% | 2 |
| Retail Services | 5.3% | 2 |
| Software Development | 7.9% | 3 |
| Telecommunications | 2.6% | 1 |
| Travel/Leisure | 2.6% | 1 |
| Other | 7.9% | 3 |

Table 2 Asset Size of Companies in the Sample

|  | Response Percent | Response Count |
|---|---|---|
| $0 - $24MM | 2.6% | 1 |
| $25MM - $99MM | 5.3% | 2 |
| $100MM - $999MM | 18.4% | 7 |
| $1 billion - $99 billion | 65.8% | 25 |
| > $100 billion | 7.9% | 3 |

### 4.1 Use of Spreadsheets after SOX

SOX emphasizes internal controls and the documentation of those internal controls. 22 out of 24 survey respondents (92%) report that spreadsheets used in the financial reporting process are of the same level or higher level of importance in the post SOX era. The number of spreadsheets that respondents reported using has not decreased as expected. 18 out of 23 respondents (78%) reported using the same number or more spreadsheets in the financial reporting process since the implementation of SOX. One might expect that after SOX, companies would either increase the controls over spreadsheets used in the financial reporting process or decrease the number of spreadsheets used in this area. However, 21 out of 24 respondents (88%) indicated they did not have a computing policy specific to spreadsheets, demonstrating lack of a comprehensive plan to mitigate spreadsheet risk. Our survey results further indicate that the operating controls over spreadsheets used in financial reporting are also still lacking in many areas including change management, version management, access control, and the development process.

### 4. 2 Self-Identified Areas of Controls Difficult to Implement

We asked the 38 respondents to identify the top three processes where implementing appropriate controls for critical spreadsheets used in financial reporting are most difficult. As shown in Figure 1, respondents were most concerned with change management, version management and access control. They were most confident with the backup process. Respondents proceeded to answer survey questions that supported their initial levels of concern or confidence in these areas as detailed in the following paragraphs. The detailed items about each area provided more insight into the perceived strengths and weaknesses of the controls. In some areas, detailed responses confirmed the strongest areas of concern. In other areas, however, later survey items suggested weaknesses in areas that were not identified by many respondents as difficult areas.

Figure 1 # of Times Process Identified as a Top 3 Area of Difficulty for Implementing Controls

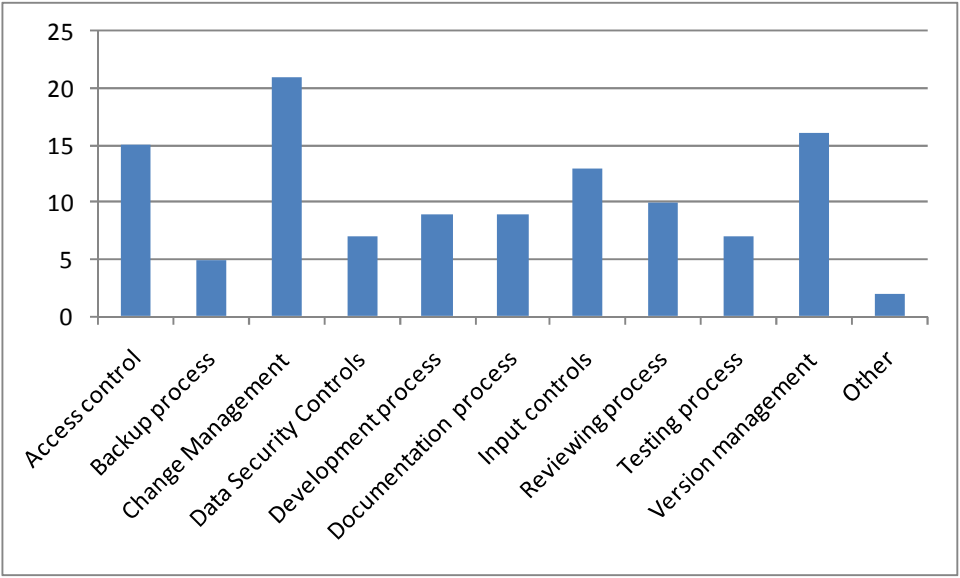

**Change Management**

Change management was the number one self-identified area of concern in implementing controls for critical spreadsheets in financial reporting as reported by 23% of the 38 respondents. Given this, it is not surprising that over one half of the 31 respondents to the questions shown in Table 3 reported no documented procedures (including use of IT and built-in spreadsheet controls) to address issues related to change management. If a company did have a documented procedure in place, the accounting department was the person or department largely responsible for ensuring the policy was followed. Two respondents listed their spreadsheet review group as the responsible department. 87% of respondents (27/31) reported they did not have a policy to describe the job title or skill requirement of the person responsible for changing the spreadsheet.

Table 3 Modifications to Spreadsheets

| Is there a documented procedure (including use of IT and built-in spreadsheet controls) to: | Yes | No |
| --- | --- | --- |
| a. Prevent unwanted changes to a spreadsheet? | 39% | 61% |
| b. Request changes to a spreadsheet? | 29% | 71% |
| c. Test accuracy of changes made to a spreadsheet? | 39% | 61% |
| d. Track changes made to a spreadsheet? | 26% | 74% |

**Version Management**

The second area respondents identified as a top concern when implementing controls over spreadsheets for financial reporting was version management. 61% of respondents (19/31) stated they did not have a documented procedure to limit access to the most recent version of the spreadsheet.

**Access Control**

Access control closely followed version management as the third area where respondents reported it was difficult to implement spreadsheet controls in financial reporting. Their answers to questions in both the development and usage phases supported this concern. 84% of respondents (26/31) stated that there was no policy to describe the job title or skill requirement of the person responsible for using a newly developed spreadsheet. Similarly when asked if they had a documented procedure to set up appropriate access levels for different identified users of the spreadsheet 71% (23/32) answered no.

**Input Control**

As shown in Figure 1, input control was an area selected by some respondents as one of their top three control concerns. However, our results indicate 61% of respondents (19/31) actually have documented procedures to validate that data inputs and outputs are complete and accurate for manual and systematic downloads. Those respondents with documented procedures in place would likely not list input controls as a top concern. For the 61% of respondents that have procedures in place, more than half of them listed the accounting department as the department or person responsible for ensuring the procedure is followed.

**Back Up Process**

While companies surveyed identified areas where it was difficult to implement controls in financial reporting, by not selecting an area we interpret this to mean they feel more confident with spreadsheet controls in that area. The backup process was the least selected area, thereby indicating the greatest level of confidence which is supported by answers to additional survey questions. 87% of respondents (27/31) reported backing up critical spreadsheets on a regular basis. 77% (23/30) reported limiting access to archived files. The majority of respondents reported the IT department was responsible for ensuring the backup procedures are followed.

**Review Process**

The review process was self-reported as a concern when implementing controls for spreadsheets used in financial reporting, but it was not in the top three concerns. Respondents' answers to subsequent survey questions supported their placement of this process as a mid-level concern. 68% of respondents (21/31) reporting no policy in place to describe the job title or skill requirements of the person responsible for reviewing the spreadsheet after new spreadsheets are developed. Over one half of the respondents (18/31) indicated the process used to review a spreadsheet included an auditing review checklist or protocol that describes the types of tests/reviews to be done. 35% (11/31) stated their review process included a procedure that generates an audit trail. For those respondents who have these review processes in place, the accounting department and internal auditor were largely identified as the department or person responsible for such review.

Further adding to the concern within the review process is the level of the reviewers' domain experience. 30% of respondents (7/23) reported a minimal level or no domain experience. The reviewers were stronger in their spreadsheet experience with 61% (14/23) reporting advanced to expert levels. If respondents utilize the review phase as a compensating control, hoping to catch possible errors made in the development stage before they would impact the financial reporting process, it is important for reviewers to have a higher level of domain knowledge than currently reported.

**Development Process**

The development process and testing process are two areas respondents did not identify as top areas of concern, selected by even less respondents than the review process. However, based on responses to later survey questions these are areas more lacking in controls than the review process. 87% of respondents (27/31) indicated there was no formal development procedure that should be followed when new spreadsheets for financial reporting are developed. 94% (29/31) stated there was no policy that described the company's styles, design and documentation standards. Furthermore, 87% of respondents (27/31) said they did not have a policy to describe the job title or skill requirements of the person responsible for developing the spreadsheet. For the 13% of respondents who did report having a policy to describe the job title or skill requirement, all reported the accounting department as the department or person responsible for ensuring the procedure is followed. The absence of formal development procedures is an issue, and this coupled with the developers' lack of accounting knowledge when creating spreadsheets critical for the financial reporting process has the potential to lead to serious errors in output. Our results indicated the spreadsheet developer's domain experience was often weak. 35% of respondents (8/23) reported minimal or no domain experience while 30% (7/23) reported only a moderate level.

**4.3 Current and Future Controls**

It is clear from the responses documented in section 4.2 that control weaknesses exist in some areas for a number of companies. We provided respondents in the longer version of the survey a list of internal controls and asked them to identify the effective controls or tools currently implemented within their company. We also asked them to identify those controls or tools they planned to implement in the future. The results are shown in Table 4.

For current effective internal controls and tools, 7 of the 15 listed were implemented in over half of the responding companies. 79% stated they have files secured in drives and server folders with limited access. This indicates respondents have taken measures to limit access to spreadsheets, but as discussed above they struggle with access controls in the development and usage phases. In addition to limiting access to the drives and folders, they also keep these areas well organized. 67% of respondents reported effective controls in logically structured directories and folders for business units, cycles, and type of spreadsheets.  Consistent with our findings, 63% of respondents report having a formal review process and 58% also report having input controls to ensure data integrity and a password to update the spreadsheet. However, independent review groups, required Excel Track Changes, external tools, developer training and a spreadsheet computing policy stating design standards are some of the controls and tools used in 25% or fewer of the responding companies.

For those respondents who stated future plans, the number one future control respondents plan to implement is Excel Track Changes followed by formal review processes.  However, the implementation percentage for any future control was less than 28% for each control.  There does not appear to be a popular control or solution that companies are eager to implement. It is particularly troublesome that both a spreadsheet computing policy for stating design standards and mandated training for developers are low in current and planned implementation.

Table 4 Internal Controls Organizations Considered for Implementation

| Internal Controls or Tools | Percent of Companies that Currently Implement Tool | Percent of Companies that Plan to Implement Tool in Future |
|---|---|---|
| Files secured in drives & server folders with limited access | 76.9% | 11.5% |
| Logically structured directories/folders for business units, cycles, and type of spreadsheets | 65.4% | 11.5% |
| Formal review process | 57.7% | 23.1% |
| Input controls that ensure data integrity | 57.7% | 15.4% |
| Password required to update spreadsheet | 57.7% | 15.4% |
| Cell protection (required) | 50.0% | 15.4% |
| More than one person responsible for data and maintenance | 46.2% | 3.8% |
| Independent review groups | 23.1% | 19.2% |
| Excel Track Changes (required) | 19.2% | 26.9% |
| Spreadsheet computing policy stating design standards | 11.5% | 15.4% |
| Mandated training for developers | 7.7% | 3.8% |
| Third party auditing software | 7.7% | 15.4% |
| Spreadsheet data consolidated into databases managed by IT | 7.7% | 11.5% |
| Third party tools for access, version, change, and archive support | 3.8% | 11.5% |
| Spreadsheet converted into server-based application | 0.0% | 11.5% |
| No Stated Plans | - | 26.9% |

### 4.4 Spreadsheet Outcomes

The emphasis on internal controls for the purposes of SOX is to increase the probability of financial statements that are materially correct. Critical weaknesses in controls over spreadsheets used for the purpose of preparing financial reports have the potential to lead to public disclosure of weaknesses in controls by external auditors. Perhaps even worse, errors in spreadsheets used in financial reporting may cause material errors in the financial statements. We asked two questions related to these possibilities. First, 46% (11/24 respondents) indicated that "internal reviews uncovered lapses or non-compliance with established protocols for spreadsheet controls." Second, 21% (5/24 respondents) answered yes to the question, "have internal reviews documented financial statement errors related to spreadsheet errors?" As mentioned earlier, compensating controls or final reviews by qualified individuals may reduce the ultimate risk of material errors in financial statements. However, the responses to these two items support the other findings that stronger internal controls are needed in many corporations.

### 5 CONCLUSIONS AND SUGGESTIONS FOR PRACTICE AND FUTURE RESEARCH

The Sarbanes-Oxley Act of 2002 has provided significant motivation for public companies to develop and tighten controls over spreadsheets used for financial reporting. There are potential serious negative consequences of poorly designed spreadsheets, including public disclosure of significant weaknesses in controls and materially misstated financial statements. Our findings demonstrate that companies continue to use spreadsheets for financial reporting. However, even with such a strong incentive for companies to have strong controls, many weaknesses in controls exist. Formal policies and procedures are still lacking in most companies for most of the stages of spreadsheets. More than half, and often most, of the companies report no policy in place to describe the required qualifications for individuals who develop, modify, review, or use spreadsheets. The results show that while individuals developing and reviewing spreadsheets have a reasonably high level of spreadsheet experience, their domain knowledge tends to be much lower. In the case of financial reporting, which can involve knowledge of complex accounting rules, this is of great concern. Though compensating controls may be in place at the final review stage before information goes into financial reports, stronger controls at earlier stages of the process would reduce the risk of non-compliance and errors.

Our findings indicate that practitioners can improve controls in several areas. More formal policies and procedures that set requirements for processes and expertise for domain knowledge and spreadsheet expertise are needed, particularly in the development, review, and use stages. We note again that the weaknesses found in this study are for controls in an area that is highly regulated and visible. We would further suggest that practitioners consider and apply similar analyses to operational spreadsheets, where errors may lead to poor business decisions.

Our findings suggest several important areas for future research. Though more research is now being done beyond the development stage, our results indicate that there are problems in all stages of a spreadsheet's life cycle. Attempts made by some organizations to control certain processes do not appear to be sufficient. Future surveys need to query organizations for more technical details about the various controls being implemented. Finally, further research comparing the impact of domain knowledge and spreadsheet expertise is also needed. The findings from this line of research will help organizations plan and implement policies that impact training for spreadsheet developers, design review, version control and auditing of spreadsheet models in all application areas within their organizations.